# Alterations of brain tissue structural complexity and disorder in Alzheimer's disease (AD): Fractal, multifractal, fractal transformation, and disorder strength analyses

Santanu Maity[1], Mousa Alrubayan[1], Mohammad Moshahid Khan[2], Prabhakar Pradhan[1]

[1]*Department of Physics and Astronomy, Mississippi State University, Mississippi State, MS, 39762*
[2]*Department of Neurology, College of Medicine, University of Tennessee Health Science Center, Memphis, TN 38163*

**Abstract**

Alzheimer's disease (AD) is characterized by progressive microstructural deterioration in brain tissue, yet conventional imaging and histopathology often lack the sensitivity needed to detect subtle early-stage changes. Here, we present a multiparametric framework combining fractal and multifractal analysis and their distributions to quantify structural alterations in human brain tissue affected by AD. Moreover, from the fractal and multifractal formalism, we introduced an innovative fractal functional distribution method, a novel technique that transforms fractal distribution into a Gaussian form. Statistically, these distribution parameters are easy to interpret and can distinguish between control and diseased tissues. Across samples, we identify pronounced threshold-dependent behavior of fractal and multifractal parameters, reflecting the intrinsic sparsity and heterogeneous intensity landscape of brain tissue. These threshold-sensitive signatures provide a framework for quantitative stage detection and may serve as biomarkers for early pathological transitions. In addition, we studied structural disorder and complexity using our established light localization technique, inverse participation ratio (IPR) analysis. IPR-based analysis demonstrates that increasing IPR pixel size highlights the elevation of structural alterations with disease progression. Together, these integrative analyses establish a robust, multi-scale quantitative framework for detecting microstructural alterations in AD, providing a promising foundation for early diagnosis and improved pathological assessment.

## 1. Introduction

Alzheimer's disease is the most common cause of dementia worldwide and poses a growing health challenge as populations age. The disease is characterized by progressive memory loss, cognitive decline, and widespread neuronal dysfunction, making early and accurate diagnosis crucial for clinical intervention and research efforts [1,2]. As AD develops and advances, the microstructural architecture of brain tissue undergoes significant changes, including alterations in cellular organization, extracellular composition, and nanoscale structural integrity. These changes, if quantified reliably and timely, offer the potential to serve as early biomarkers of neurodegeneration [3]. The current diagnostic approach relies on histopathological examination of biopsy or postmortem brain tissue samples with manual staining and interpretation. However, this method is more subjective and depends on human expertise, and it is not accurate in detecting structural alterations at the nanoscale [4,5]. To overcome these limitations, recent research has increasingly turned to bright-field transmission microscopy, which enables high-resolution visualization of stained tissue sections and can be paired with quantitative computational analysis to enhance diagnostic precision [6,7]. Automated systems integrating bright field transmission imaging with advanced analytical frameworks have shown potential for improving disease detection, staging, and reproducibility across clinical and pathological workflows [6,8–10].

Biological tissue samples are intrinsically heterogeneous systems with increasing structural complexity and disorders as disease progresses, which are linked to mass density fluctuations and result in refractive index variations [6,9,11]. Visualizing these structural changes remains challenging, prompting the application of mathematical tools to characterize tissue organization across scales. In this context, fractal and multifractal analyses offer powerful descriptors of self-similarity, heterogeneity, and irregularity in biological structure [6,8,11]. Fractal dimension serves as a potential biomarker for disease diagnosis [12]. However, grayscale threshold intensity plays a vital role in image binarization. We studied the threshold dependence behavior of fractal dimension and identified a novel trajectory that can distinguish control samples from AD disease samples. We also obtained the optimal threshold at which maximum changes in fractal dimension occurred for these two groups [9].

To strengthen statistical interpretation, we further characterized tissue complexity using fractal and multifractal distribution-based measures. Fractal dimension values derived from long-tailed distributions showed clear shifts in mean and variance from control to AD tissue, while multifractal spectra captured subtle region-dependent variations across scales.

Building on these results, we introduced a fractal functional transformation, a new computational method that converts fractal dimension distributions into an approximately Gaussian form, yielding intuitive distribution parameters suitable for statistical analysis [8,9]. Finally, we apply inverse participation ratio (IPR) analysis, a physics-based method for light localization, to directly assess nanoscale structural disorder from the optical transmission intensity of tissue samples [9,11].

Together, this multiparametric quantification approach for studying structural alterations in AD brain tissue establishes statistically significant, quantifiable, and interpretable frameworks for disease diagnosis. Obtaining several quantification parameters from the analysis provides an easy pathway to detect the disease at its earliest stage, improve staging of disease progression, and enhance clinical pathology evaluation. Our study combined all these methods into a framework that includes fractal analysis, multifractal analysis, threshold dependence behavior of fractal dimension, fractal functional transformation, and IPR analysis to enhance the diagnosis and evaluation of AD brain disease.

## 2. Methods and Results

### 2.1 Sample Information: Brain TMA

#### 2.1.1. Sample Acquisition

Brain samples are collected from BioChain Institute Inc., including specimens from Alzheimer's Disease (AD) patients and their control counterparts. Five slides from AD cases were purchased from BioChain, featuring various tissue samples from the precentral gyrus, postcentral gyrus, occipital lobe, and cerebellum. Each slide has eight cores for both control and AD brain tissues. The samples are 5 μm thick and placed on glass slides. Tissue microarray samples are acquired from the precentral gyrus of the brain for AD-related studies.

#### 2.1.3. Instrumentation for Automated Scanning of TMA Images: Optical Transmission Microscopy

Imaging of biological samples was performed using an Olympus BX61 bright-field optical transmission microscope (Tokyo, Japan), and images were processed with Amscope software. In our earlier research on cancer detection, we showed that a change in refractive index causes a linear increase in the sample's mass density. As the disease progresses, the small linear mass density increases due to filling of the tissue porosity. This change results in variations in refractive index, which correlates transmission pixel intensity to the sample's mass density [6]. The following equations illustrate the relationship between transmission intensity and mass density.

$$I_t \propto (n_{tissue+glass} - n_{air}) \propto changes\ in\ tissue\ mass\ density.$$

Here, $n_{tissue+glass}$ is the total refractive index of tissue mounted on a glass slide and $n_{air}$ is the refractive index of the air. The above equation shows the direct link between transmission intensity and variations in refractive index and mass density. Fractal dimension is associated with variations in the mass density of tissue samples, so analyzing the transmission intensity from the sample provides a way to estimate the fractal dimensions, which can be used as a biomarker for disease detection.

### 2.2. Fractal Dimension Analysis of Brain TMA Sample

#### 2.2.1. Box-Counting Method for Estimating Fractal Dimension

The box counting method is the primary approach for calculating the fractal dimension of a 2D binary image. Using this method, we calculated the fractal dimension of biological tissue samples. The process involved converting grayscale images of biological tissue samples into 2D binary images and then applying the standard box-counting formula to calculate the fractal dimension. The following formula for calculating fractal dimension is described below. [6]

$$D_f = \frac{ln(N(r))}{ln\left(\frac{1}{r}\right)}$$

Here $D_f$ is the fractal dimension of the structure, and N(r) represents the number of non-empty boxes that cover the image at box size r.

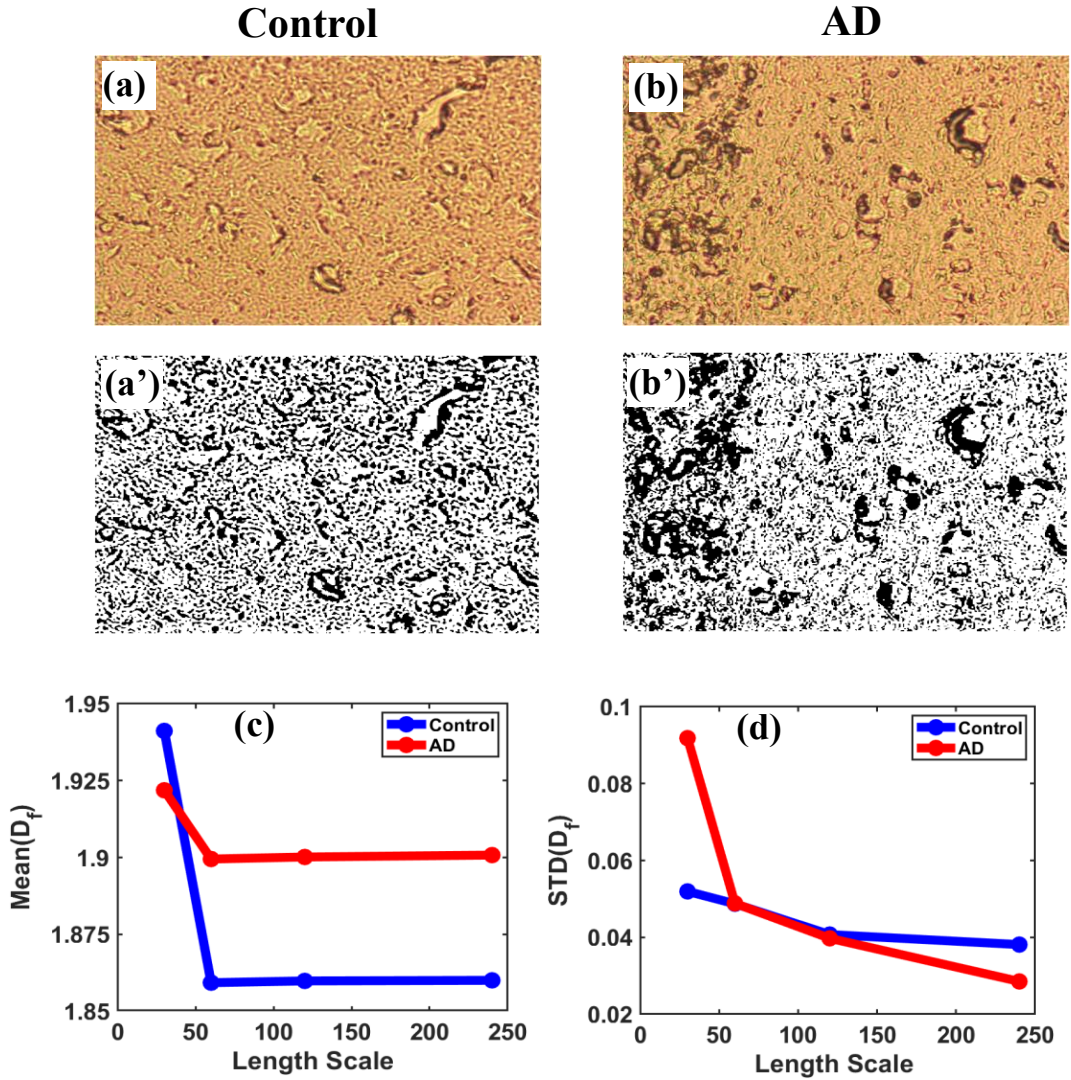

**Figure 1:** **(a)** and **(b)** show brightfield images of control and AD brain tissues, respectively, while their corresponding binary images are displayed in **(a')** and **(b')**.
Variation of local fractal dimension ($D_f$) in brain tissue images; **(c)** its mean and **(d)** standard deviation as a function of length scale (region size), comparing control and Alzheimer's disease (AD) groups.

The local fractal dimensions were determined using the box-counting method, with box sizes ranging from 30×30 to 240×240 pixels applied to image regions of various sizes (30×30, 60×60, 120×120, and 240×240 pixels). By averaging the slopes across different scales, the mean and standard deviations of the fractal dimension were quantified. The consistent pattern across regions of varying sizes highlights the scale-invariant property of the fractal dimension in differentiating between control and AD brain tissues.

### 2.2.2. Threshold Dependent Behavior of Grayscale intensity variation: Binary filling in fractal dimension calculation

From our previous study [9], We discovered that the threshold emerges as a prominent biomarker for distinguishing stages during the binary filling in fractal dimension analysis. Our previous studies [6,10,13], highlighted the importance of selecting a default threshold during grayscale-to-binary image conversion for fractal dimension estimation. Grayscale pixel intensities below 50% of the maximum grayscale value are assigned a value of 0, while those above 50% are assigned a value of 1. However, our recent study [9], explored the threshold further and found threshold-dependent behavior in the fractal dimension. We identified the optimal changes in fractal parameters at a specific threshold, and the threshold region beyond which the control and disease groups are separated.

Due to the sparsity and more complex structure of brain samples, we investigated how the fractal dimension varies with threshold during binary image filling for fractal dimension estimation. The relationship between variation in threshold and fractal dimension, shown by a novel trajectory, indicates that the mean fractal dimension, *Mean($D_f$)*, and its standard deviation, *STD($D_f$)*, vary with threshold values. At an optimal threshold value, the maximum fractal dimension values are observed for the control and AD groups, emphasizing the maximum differences between the two groups for disease diagnosis.

### 2.2.3. Fractal Parameters Mean($D_f$) and STD($D_f$) and their variation with grayscale threshold intensity

Figure 2 shows the variation in mean fractal dimension (*Mean($D_f$)*) and its standard deviation (*STD($D_f$)*) across 10 micrographs per group, demonstrating fractal dimension variation as a function of threshold pixel percentage.

The trajectory below identified threshold values exceeding 42.968%, corresponding to a grayscale value of 110 on the grayscale threshold intensity scale from 0 to 255, indicating a noticeable difference between the control and AD groups.

Figure 2(a) shows the maximum difference in mean fractal dimension between the control and AD groups, which occurs at an optimal threshold of 60.546%. The *Mean ($D_f$)* values for both the control and AD disease groups are 1.8593 and 1.8995, respectively, at this optimal threshold.

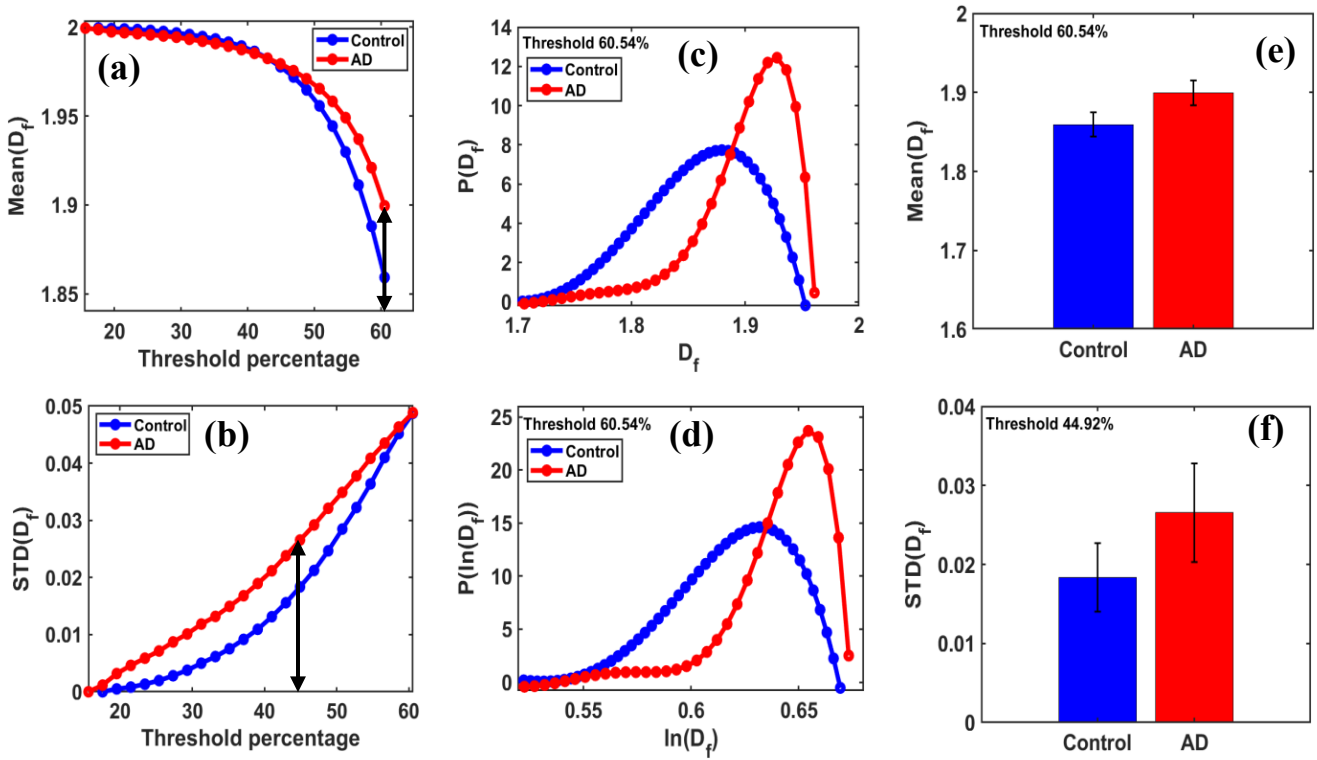

Figure 2 (a) shows the mean fractal dimension (*Mean(D$_f$)*) and (b) the standard deviation of *D$_f$* (*STD(D$_f$)*) as functions of threshold pixel percentage, ranging from 10% to 60.54% of the grayscale, with the default threshold set at 50%. It highlights a clear peak trend between the control and AD groups, and identifies an optimal threshold at which the maximum differences in these parameters occur. For this study, binary regions of size 60 × 60 pixels were used for box-counting fractal dimension analysis, with box sizes set to [2, 4, 6, 10, 12].

Figure 2 (c) shows the fractal distribution *P(D$_f$)* versus *D$_f$* plot, which has a slightly extended tail. (d) The *P(ln(D$_f$))* versus *ln(D$_f$)* plot shows a more extended non-Gaussian distribution. The most noticeable difference between the control and AD groups in the mean values of the *D$_f$* curves occurs at a threshold of approximately 60.54%.

Figure 2 (e) displays the bar graphs of the *Mean(D$_f$)* for control and AD groups, where the *Mean(D$_f$)* value is higher, showing a 2.164% increase in the AD group compared to the control group. (f) The bar graphs of the standard deviation (STD) for the control and AD groups indicate that the *STD(D$_f$)* value is higher, with a 44.654% increase in the AD group compared to the control group.

Figure 2(b) shows the maximum difference in *STD(D$_f$)* between the control and AD groups at an optimal threshold of 44.921%. The standard deviations of *D$_f$* (*STD(D$_f$)*) for both the control and AD groups are 0.0184 and 0.0266, respectively, at this optimal threshold.

Figure 2(c) shows the distribution of the fractal dimension, which exhibits a slightly extended one-tailed distribution. Figure 2(d) shows the logarithm of the fractal dimension, revealing a more long-tailed distribution that appears less Gaussian. These distributions are generated at the optimal threshold value for control and AD groups, capturing the fractal dimension as a function of grayscale threshold intensity variation between the two groups.

Figure 2(e) shows the bar graphs of the *Mean*($D_f$) for control and AD groups, where the *Mean*($D_f$) value is higher, showing a 2.164% increase in the AD group compared to the control group. Figure 2(f) The bar graphs of the standard deviation (STD) for the control and AD groups indicate that the *STD* ($D_f$) value is higher, with a 44.654% increase in the AD group compared to the control group.

## 2.2.4. Fractal analysis of grayscale images to identify key regions and determine optimal parameters for disease diagnosis.

Fractal and multifractal analysis of various regions in 2D grayscale images were performed to show how the fractal dimension changes with different thresholds and sample sizes. First, the 2D grayscale images were converted into binary format by applying different threshold values and then divided into regions of various sizes (30×30, 60×60, 120×120, and 240×240 pixels). We analyzed variations in fractal dimension across these binary regions for a range of grayscale threshold intensities (25-155), demonstrating changes in fractal dimension at different threshold levels. For control and AD brain tissues, we observed threshold-dependent behavior in the fractal dimension, aiming to identify the optimal threshold values at which the difference in fractal dimension between control and AD brain tissues is maximal.

The box-counting method with box sizes [2, 4, 6, 10, 12] was applied across regions to capture fine structural variations at multiple spatial scales during fractal dimension estimation. The analysis was conducted on images 480×480 pixels in size, with a smaller 60×60-pixel region. For example, in regions such as (1,4) and (1,5), which refer to specific coordinates within the brain tissue image, exhibited a systematic increase in the fractal parameter *Mean*($D_f$) from control to AD tissues. The primary objective was to identify the maximum changes in *Mean*($D_f$) between these two groups within smaller, specific regions of tissue structure.

One key aspect of the biological samples used in this study is that they are paraffin-embedded to preserve the overall structure of the tissue samples. Still, this method subtly masks some pathological changes in the image structure when observed at the macroscopic scale. Nevertheless, mass accumulation during disease progression means this mass localization may be present, and it can be illustrated and emphasized through detailed image analysis. Our binary representation of the biological tissue structure reveals some discrete regions in the tissue that hold structural abnormalities or heterogeneity due to mass accumulation in that region, such as those at specific coordinates like (1,4), (1,5), (1,8), (2,4), and (3,6), which show significant structural differences, suggesting potential pathological importance.

These findings underscore the importance of examining localized features within the tissue microenvironment, even when global structural changes appear minimal.

To support this, we applied the Inverse Participation Ratio (IPR) to study the structural disorder at the nanoscale, which varies with length scale in control and AD disease tissue, offering more profound insights into the complexity and heterogeneity of tissue structures.

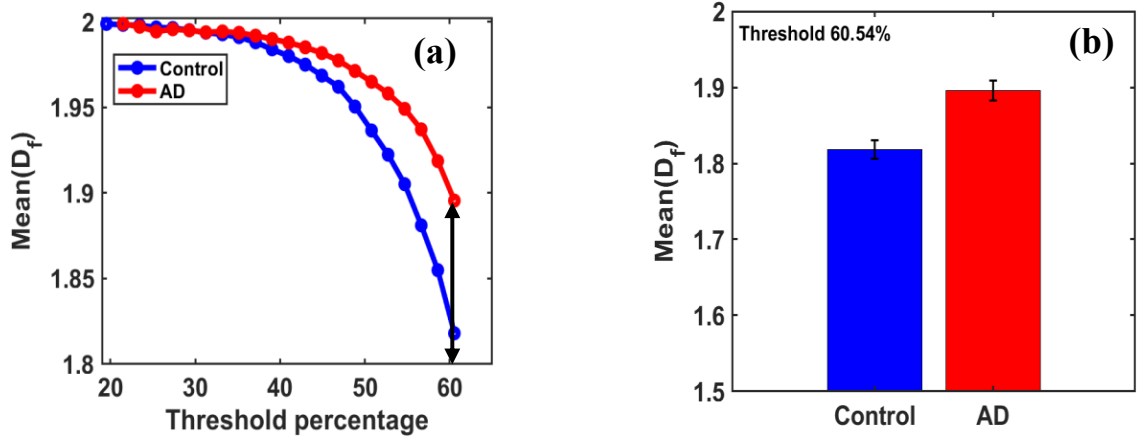

Figure 3. (a) For the binary region at coordinates (1, 5), with a size 60 × 60 pixels used for box-counting fractal dimension analysis with box sizes = [2, 4, 6, 10, 12] highlights the variation of mean fractal dimension ($Mean(D_f)$) across grayscale thresholds ranging from 10% to 60.54% (default threshold at 50%). It displays distinct peak patterns that vary between the control and AD groups. From this trajectory, an optimal threshold value was identified, at which the maximum difference in Mean fractal dimension ($Mean(D_f)$) between the two groups was observed.

Figure 3. (b) displays a bar graph of the $Mean(D_f)$ for the control and AD groups, showing that the mean fractal dimension ($Mean(D_f)$) is higher in the AD group than in the control group, with a percentage increase of 4.266%.

Figure 3(a) shows the threshold-dependent behavior of the fractal dimension, and for threshold values exceeding 31.25%, corresponding to a grayscale value of 80, a clear distinction between the control and AD groups is observed. The maximum difference in the mean fractal dimension between these two groups occurs at an optimal threshold of 60.546%. At this threshold, the $Mean(D_f)$ values for the control and disease samples are 1.8180 and 1.8956.

Figure 3(b) shows bar graphs of fractal dimension for the control and AD groups, indicating that the mean fractal dimension($Mean(D_f)$) is higher in the AD group than in the control group.

Our preliminary studies aimed to identify key regions and the optimal fractal parameter ($Mean(D_f)$) that can distinguish between the control and AD stages, based on the systematic increase in fractal dimension with disease progression. However, further studies are required to obtain more conclusive findings and results.

## 2.3. Multifractal analysis and formalism: pixel intensity distribution of grayscale images and their multifractal spectrum via f(α) vs. α plot

The traditional box-counting method for fractal analysis can assess the structural complexity related to disease progression. However, brain samples are a sparser medium, exhibit more complex structures, and display multifractal characteristics. In this study, spectrum scaling behavior in multifractal analysis was used to examine how structural complexity varies across regions of tissue structure.

We divided the 2D brain tissue sample into L×L boxes of side length ε. Then we showed the pixel intensity distribution in tissue structure; the probability distribution of the pixel intensities at the $i^{th}$ cell is expressed as [8,9,14,15]

$$P_{\varepsilon,i} = N_\varepsilon(i)/N_{total} \sim \varepsilon^{\alpha_i}$$

$P_{\varepsilon,i}$ is the probability of pixel intensity distribution at each box, defined as the number of pixel intensities in the ith box $N_\varepsilon(i)$ over the total number of pixel intensities $N_{total}$ in the tissue structure.

The equation for the multifractal spectrum is derived as follows: a detailed description can be found elsewhere. [8,14]

$$\mu_{i(Q,\varepsilon)} = P_{i(Q,\varepsilon)}{}^Q / \sum_{i=1}^{n\varepsilon} P_{i(Q,\varepsilon)}{}^Q$$

We will get the spectrum equation:

$$f(\alpha_Q) = Q \times \alpha_Q - \tau_Q = \sum_{i=1}^{n\varepsilon} (\mu_{i(Q,\varepsilon)} \times ln\,(\mu_{i(Q,\varepsilon)}))/\,ln\,\varepsilon.$$

Here, the Q value ranges from -10 to 10. The multifractal spectrum *f(α)* represents the distribution of the fractal dimension *α*, which reflects mass density fluctuations across different regions of the tissue caused by inherent sparsity and heterogeneity.

Figure 4 demonstrates a broader spectrum in the Alzheimer's disease group; the multifractal spectrum illustrates more heterogeneity and structural complexity that increase with disease progression. However, more uniform structural behavior was observed across the control tissue samples, resulting in a narrower spectrum. [16]

According to the multifractal theorem [17,18], AD tissue displays a multifractal nature due to irregular accumulation of masses and structural complexity, resulting in a non-uniform distribution of masses within the tissue. However, control groups show a uniform distribution of masses.

For parametrization to quantify disease from the control group, some spectrum parameters are extracted from the multifractal spectra, such as bandwidth ($\Delta\alpha$), spectrum height ($\Delta f$), maximum singularity exponent ($\alpha_{max}$), minimum singularity exponent ($\alpha_{min}$), and multifractal strength parameters, which include peak singularity strength ($f_{max}$) and minimum singularity strength ($f_{min}$).

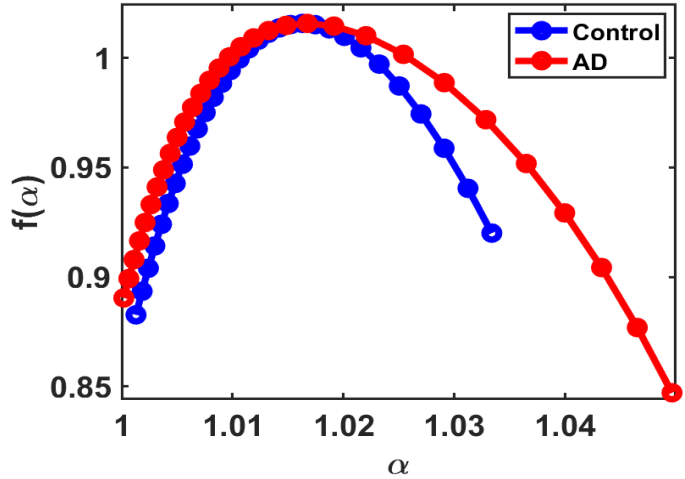

Figure **4** shows the Multifractal spectrum *f(α)* versus *α* plot for the control and AD groups, each with a sample size of N = 10. Minor changes are observed in the spectral distribution between control and disease. (Average Multifractal Spectrum).

| Parameter | Control | AD |
|---|---|---|
| $\alpha_{min}$ | 1.0013 | 1.0002 |
| $\alpha_{max}$ | 1.0334 | 1.0497 |
| $\Delta\alpha$ | 0.0321 | 0.0495 |
| $f_{min}$ | 0.862171 | 0.823419 |
| $f_{max}$ | 1.015763 | 1.015763 |
| $\Delta f$ | 0.1536 | 0.1923 |

The results obtained from the spectrum distribution align with previous findings [9,19], showing that the spectrum's bandwidth ($\Delta\alpha$) and spectrum height ($\Delta f$) increased from the control to the AD group as the disease progressed, indicating greater structural complexity, disorder, and heterogeneity in disease groups. However, this method is challenging for quantification because differences in the multifractal parameters between the control and AD groups are relatively small. In contrast, visual inspection of the spectrum distribution can enhance the possibility of detecting stages and be useful for further diagnosis.

## 2.4. Introducing a new approach, fractal functional transformation

### 2.4.1. Obtain the Gaussian distribution process through the functional transformation of the fractal dimension.

Our recent study on improving disease diagnosis by introducing a new theoretical concept: the functional transformation of fractal dimension at each point in the 2D tissue micrograph. We observed that the distribution of fractal dimensions is long-tailed. Still, our goal was to obtain a Gaussian distribution, which is easier to interpret statistically, and the parameters derived from it served as quantification parameters.

Further implementation included the functional transformation of fractal dimensions, described below [8,9].

$$D_{tf} = \frac{D_f}{D_{fmax} - D_f}$$

Here $D_{tf}$ demonstrates a functional transformation of fractal dimension at each point and $D_{fmax} = 2$ for 2D tissue micrograph.

The distribution of the fractal functional transformation is broader and exhibits longer tails. However, its lognormal distribution lies within the Gaussian space, and it becomes Gaussian. The distribution parameters, such as Mean and Standard Deviation (STD) of $ln(D_{tf})$, serve as quantification parameters for the diagnosis process. This method represents a significant improvement in diagnostic methods or disease detection and serves as a potential biomarker.

Our multiparametric framework included fractal analysis, multifractal analysis, threshold-dependent fractal-dimension behavior, fractal functional transformation, and inverse participation ratio, making it a highly impactful method for studying structural changes in biological samples for disease detection.

### 2.4.2. Threshold-dependent nature of grayscale intensity variation for binary filling in fractal dimension calculations and its effects on fractal functional transform parameters.

We observed variation in the fractal dimension with threshold within the grayscale intensity range of 0-255. During binary filling in the fractal dimension calculation, the threshold significantly influences the fractal dimension's variation. Similarly, the fractal functional transformation parameters also vary with grayscale threshold intensity, exhibiting a novel trajectory beyond a specific range, where a clear separation between control and disease occurred. From this trajectory, a relationship was established showing changes in fractal dimension related to grayscale threshold intensity, which is correlated with mass density in pixel intensity values of a 2D fractal.

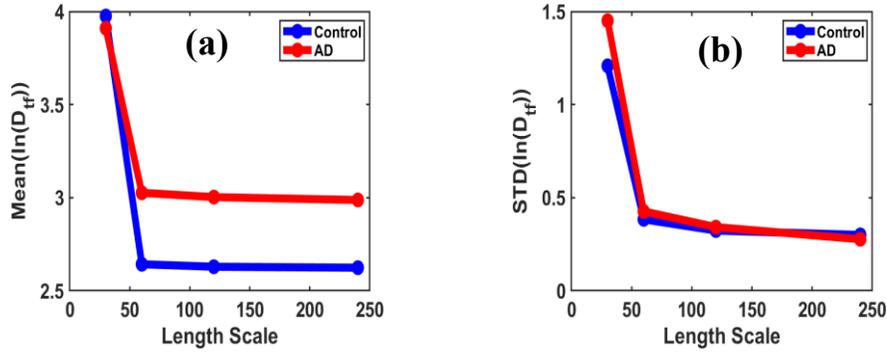

Figure **5.** Variation of *ln(Dtf)* parameters in brain tissue images; **(a)** its mean and **(b)** standard deviation as a function of length scale (region size), comparing control and Alzheimer's disease (AD) groups.

Figure 5 illustrates the scale-variant nature of fractal-transformed parameters between the control and AD groups. The box counting method for calculating fractal dimension was applied to smaller regions of the image at different length scales. Initially, the images (480×480 pixels) were divided into smaller areas (30×30, 60×60, 120×120, and 240×240 pixels), and their functional distribution parameters, *Mean*($ln(D_{tf})$) and *STD*($ln(D_{tf})$), were further analyzed from the fractal dimension.

**2.4.3. Distribution of the transformed fractal on a logarithmic scale (ln(Dtf)) becomes Gaussian.**

Threshold-dependent behavior of fractal functional transform parameters, *Mean*($ln(D_{tf})$) and *STD*($ln(D_{tf})$), reveals a new trajectory that identifies the optimal threshold at which the maximum differences in these quantification parameters occur between the control and AD disease groups. The distribution of these parameters becomes Gaussian and statistically interpretable. We observed that the distribution parameters, *Mean($ln(D_{tf})$)* and *STD($ln(D_{tf})$)*, increased with disease progression, especially at higher threshold values. Therefore, these parameters can be used as quantification metrics to distinguish between the control and AD groups.

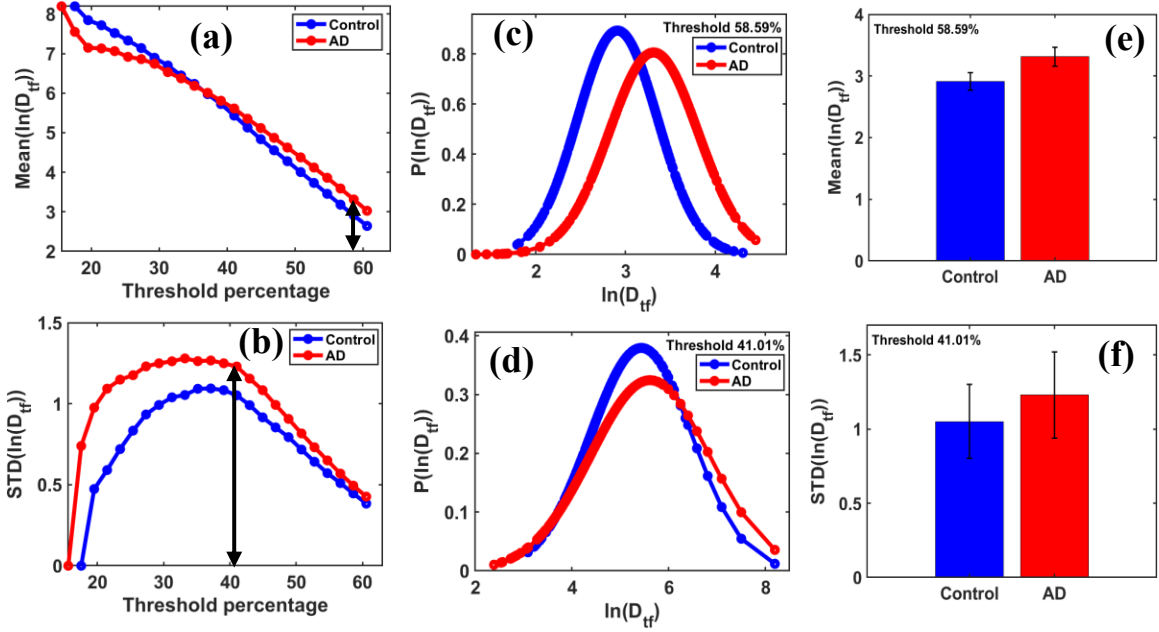

Figure **6** shows **(a)** *Mean(ln(D_tf)* and **(b)** *STD(ln(D_tf))* as functions of threshold pixel percentage, ranging from 10% to 60.54% of the grayscale, with the default threshold set at 50%. It highlights a clear peak trend between control and AD disease groups, and an optimal threshold value is identified where the maximum differences in these parameters occur. For this study, binary regions of size 60 × 60 pixels were used for box-counting fractal dimension analysis, with box sizes set to [2, 4, 6, 10, 12].

**(c)**: The distribution of *P(ln(D_tf))* versus *ln(D_tf)* follows a Gaussian distribution based on the best polynomial fits of the brain sample, with a chi-square test for Gaussian fitting showing scores above 90%. The most significant difference in the *Mean*(ln(Dtf)) values between the control and AD groups was observed at a threshold of 58.59%.

**(d)** The distribution of *P(ln(D_tf))* versus *ln(D_tf)* follows a Gaussian distribution based on the best polynomial fits of the brain sample, with a chi-square test for Gaussian fitting yielding scores over 90%. The most significant difference in the standard deviation of *ln(D_tf)* between control and AD groups was observed at a threshold value of 41.01%.

Figure **6 (e)** shows the bar graphs of the *Mean(ln(D_tf))* for control and AD groups, with the AD group having a 13.382% higher mean, indicating an increase. **(f)** The bar graphs show the standard deviation of ln(Dtf) (*STD(ln(Dtf))*) for the control and AD groups; the *STD(ln(Dtf))* is higher in the AD group by 16.931%, indicating greater variability.

From the above trajectory, Figure 6 identified threshold values exceeding 37.1094%, corresponding to a grayscale value of 95 on the grayscale threshold intensity scale from 0 to 255, indicating a noticeable difference between the control and AD groups.

Figure 6(a) shows the maximum difference in *Mean(ln(Dtf))* values between the control and AD groups, occurring at an optimal threshold of 58.59%.

Figure 6(b) displays the maximum difference in $STD(ln(D_{tf}))$ between the control and AD groups at an optimal threshold of 41.0156%.

The probability distribution of $ln((D_{tf}))$ is associated with maximum mean values shown in Figure 6(c), and the probability distribution of $ln(D_{tf})$ corresponding to maximum standard deviations presented in Figure 6(d). These findings indicate that the lognormal transformation of functional parameters results in Gaussian distributions, thereby providing a new diagnostic framework.

Figure 6 (e) and (f) display the fractal functional parameter values at the optimal threshold values using bar graphs for two groups.

The *Mean ($ln(D_{tf})$)* values for control and AD groups at the optimal threshold of 58.59% of maximum grayscale threshold intensity are 2.9098 and 3.3123, respectively, as shown in Figure 6(e).

The *STD($ln(D_{tf})$)* values for control and AD groups at the optimal threshold of 41.01% of maximum grayscale threshold intensity are 1.0518 and 1.2298, respectively, as shown in Figure 6(f).

These quantification parameters further improve disease diagnosis and are statistically very easy to interpret from the distribution plots.

## 2.4.4. Binary fractal dimension through different length scale analysis: Variation in grayscale threshold during fractal dimension calculation and its effects on functionally transformed parameters and their distribution.

We previously demonstrated binary fractal dimension analysis across different length scales in 2D tissue micrographs. We observed that applying the box-counting method to the tissue micrographs for the region at coordinates (1,5) involved converting the image to binary, with a size of 60×60 pixels, and computing its fractal dimension using box sizes [2, 4, 6, 10, 12]. The average fractal dimension increased by 4.266% from the control to the AD group.

Additionally, the fractal functional transformation parameter $Mean(ln(D_{tf}))$ varies with grayscale threshold intensity across the scale of 0-255.

As shown in Figure 7 (a), a threshold value of 50.781% indicates the point where the most significant separation in the peak values of the $ln(D_{tf})$ occurs between the control and AD groups.

The probability distribution of $ln(D_{tf})$ relates to the maximum mean values shown in Figure 7(b). The $Mean(ln(D_{tf}))$ values for the control and AD disease groups at the optimal threshold of 50.78% of maximum grayscale threshold intensity are 3.5135 and 4.2606, respectively, as shown in Figure 7(c).

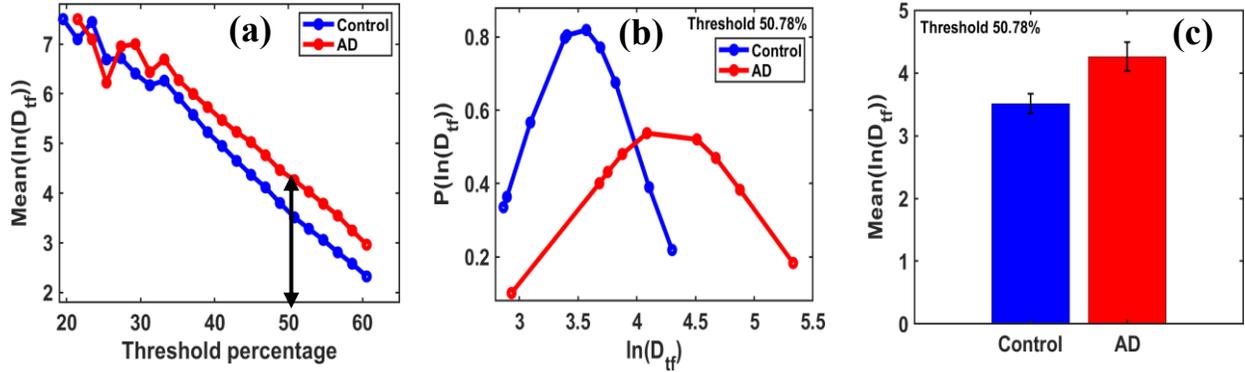

Figure 7 **(a)** For the region at coordinates (1, 5), with a size of 60×60 pixels, used for box-counting fractal dimension analysis with box sizes = [2, 4, 6, 10, 12], highlights the variation of *Mean(ln($D_{tf}$))* across grayscale thresholds ranging from 10% to 60.54% (default threshold at 50%). It displays distinct peak patterns that vary between the control and AD groups. From this trajectory, an optimal threshold value was identified, at which the maximum difference in *Mean(ln($D_{tf}$))* between the two groups was observed.
**(b)**: The distribution of *P(ln(Dtf))* versus *ln($D_{tf}$)* follows a Gaussian distribution based on the best polynomial fits of the brain tissue samples, with a chi-square test for Gaussian fitting showing scores above 90%. The most significant difference in the *Mean(ln($D_{tf}$))* values between the control and AD groups was observed at a threshold of 50.78%.
**(c)** The bar graphs of the *Mean(ln($D_{tf}$))* for control and AD groups, with the AD group having a 21.264% higher mean, indicating an increase.

## 2.5. Inverse Participation Ratio (IPR) Analysis: Studying structural disorder via light localization of transmission intensity

The Inverse Participation Ratio (IPR) analysis is a physics-based method used to measure light localization and assess nanoscale structural disorders in heterogeneous dielectric media, such as biological tissues. Biological tissues are not uniform; their mass density and refractive index vary at the nanoscale. In brain samples, as the disease progresses, fluctuations in mass density increase due to the rearrangement of DNA, RNA, lipids, and proteins, leading to non-uniform mass accumulation. This causes variations in the refractive index, which are linearly dependent on mass density fluctuations. In uniform media, such as control samples, light spreads evenly through the tissue structure, resulting in more minor fluctuations. Conversely, in disordered media such as tissues affected by Alzheimer's disease, fluctuations are random, leading to a greater localization of photons. To quantify the degree of structural disorder, IPR analysis is used for statistical comparison by examining the mean and standard deviation of IPR values between control and diseased tissues. Control brain samples typically have low IPR values, indicating lower structural disorder and less light localization, while higher IPR values in diseased tissues indicate greater structural disorder [9,11,20–23].

A direct relationship is established between change in mass density $\rho(x,y)$ in a voxel of cells and refractive index variation; they are linearly dependent. Additionally, the transmission intensity resulting from scattering is also linearly proportional to mass density variation. It is represented in the following equation.

$$I(x,y) \propto \rho(x,y) \propto n(x,y)$$

Here $I(x,y)$ represents the transmission intensity from the sample and $n(x,y)$ is the change in refractive index.

Refractive index fluctuations create an optical lattice potential $\varepsilon_i$, which can be represented as

$$\varepsilon_i = \frac{dn(x,y)}{n_0} \propto \frac{dI(x,y)}{I_0}$$

Anderson's tight-binding model (TBM) demonstrates that light can be trapped in disordered cells, just as electrons are localized in random lattices. Here, biological tissues act as optical lattices [24–26].

The tight-binding lattice model describes the Hamiltonian of a disordered system. The Hamiltonian of the optical lattice is defined as

$$H = \sum \varepsilon_i |i><i| + t \sum \langle ij \rangle \, (|i><j| + |j><i|)$$

Here, $\varepsilon_i$ is the optical potential energy of eigenfunctions at the $i^{th}$ lattice site. $\langle ij \rangle$ representing the nearest-neighbor hopping interaction between these two lattice sites, and t is the overlap integral. $|i>$ and $|j>$ are the optical eigenfunctions at the $i^{th}$ and $j^{th}$ lattice sites.

In IPR analysis, mean IPR is calculated using the following equation,

$$\langle IPR \rangle_{L \times L} = \frac{1}{N} \sum_{i=1}^{N} \int_0^L \int_0^L E_i^4(x,y) dx dy$$

In the above equation, $E_i$ represents the optical eigenfunctions of the Hamiltonian at the $i^{th}$ lattice site. Mean IPR quantifies the degree of structural disorder. N defines the total number of optical eigenfunctions in the lattice of size L×L.

IPR value is linearly proportional to the refractive index fluctuations $<dn>$ and correlation length $l_c$. The following equation is defined as follows:

$$\langle IPR \rangle_{L \times L} \sim <dn>.l_c$$

On the other hand, IPR quantifies the structural disorder parameter, $L_{d-ipr}$, which also depends linearly on it.

After summing up all these equations above, we derive the following relation.

$$\langle IPR \rangle_{L \times L} \propto L_{d-ipr} \sim <dn>.l_c$$

$$STD(IPR)_{L \times L} \propto L_{d-ipr} \sim <dn>.l_c$$

These equations provide a robust method for estimating IPR parameters, enabling easy detection of structural disorder during disease progression by calculating the mean and standard deviation of IPR.

### 2.5.1. IPR Analysis of Brain TMA Samples and Findings

Figure 8 displays the grayscale images and their corresponding IPR images for control and AD disease groups. For IPR analysis, the 10 best images were selected from each group. The brightfield images of control brain tissue samples are shown in Figure 8 ($a_1$) and ($a_2$), with their corresponding IPR images in Figures ($a_1$') and ($a_2$'). For Alzheimer's disease, brightfield images shown in figures 8(b1) and 8 (b2) have corresponding IPR colormap images highlighted in figures 8 (b1') and 8 ($b_2$').

In the figure below, Figures 8 (c) and (d) show that IPR parameters increase with increasing pixel size. Figure 8(c) demonstrates that the mean IPR rises as the pixel size of optical eigenfunctions grows. Similarly, Figure 8(d) shows that the standard deviation of IPR *(*STD of IPR) also increases with larger pixel sizes. The IPR pixel size plays a direct role in IPR variation. Larger eigenfunctions in the optical lattices imply that each eigenfunction has a different mass density distribution. Therefore, as the eigenfunction pixel size increases, structural disorder also increases due to greater mass density fluctuations, and IPR values also rise systematically.

The bar graphs below show that IPR parameters are higher in AD disease samples than in control brain samples. Figure 8(e) shows that the Mean IPR increased by 13.32% from the control group to the AD group, and Figure 8 (f) shows that the STD of IPR increased by 69.34% in the AD group relative to the control. Higher IPR values indicate greater structural disorder at the nanoscale in brain tissue samples as the disease progresses. This method is very important and accurate for diagnosing the disease at its earliest stage.

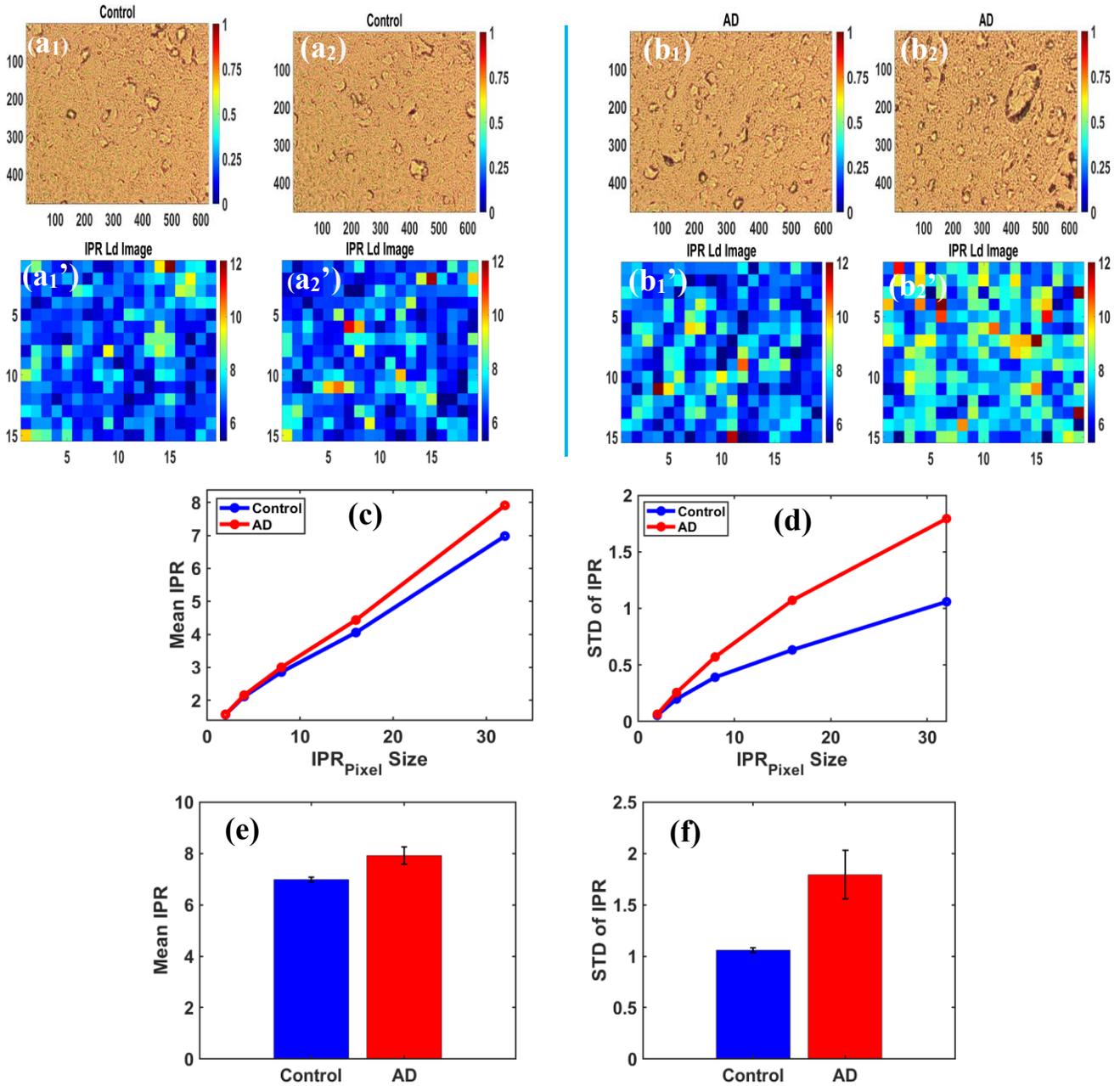

Figure 8. (a₁) and (a₂) display brightfield images of healthy control brain tissues. Meanwhile, (a₁') and (a₂') show the corresponding IPR colormap images of the control brain tissues. (b₁) and (b₂) display brightfield images of Alzheimer's disease tissues, and (b₁') and (b₂') show the corresponding IPR colormap images of the Alzheimer's disease tissues. Changes in mean IPR values are linearly related to structural changes in tissue, as demonstrated by $<IPR> \sim L_d$-IPR color map.
(c) Mean IPR plotted against IPR pixel size for the Control and AD groups. The observed percentage changes in Mean IPR between the groups were 5%, 9.36%, and 13.32% for matrix sizes of 8, 16, and 32, respectively. (d) STD of IPR versus IPR pixel size for the Control and AD groups. The observed percentage changes in Mean IPR between the groups were 17.49%, 29.29%, 45.84%, 69.28%, and 69.34% for matrix sizes of 2, 4, 8, 16, and 32, respectively.

Bar graphs showing **(e)** mean IPR and **(f)** standard deviation (STD) of IPR in control and AD tissues. IPR parameters are significantly higher in AD, with the mean IPR being 13.32% higher and the STD of IPR being 69.34% higher in AD tissues compared to control tissues.

## 3. Discussion and Conclusions:

We presented a multiparametric quantitative framework to characterize structural alterations in AD brain tissues using bright-field images of tissue microarrays. By integrating fractal analysis, multifractal characterization, a fractal functional transformation, and Inverse Participation Ratio (IPR)-based light localization, we provide a unified approach for probing tissue complexity across multiple spatial scales. This framework yields a suite of quantifiable parameters that capture disease-associated structural changes not readily discernible by conventional histopathology. Our findings resonate with and extend a growing body of literature that uses complexity measures to probe disease-associated structural changes.

Several neuroimaging studies have shown that fractal dimension ($D_f$) is sensitive to the loss of cortical complexity in AD and related conditions. For example, King and coworkers reported that $D_f$ of the cortical ribbon derived from MRI distinguishes AD from controls and correlates with cognitive deficits [12]. More recently, Marzi et al. demonstrated that cortical gray-matter $D_f$ predicts conversion from mild cognitive impairment to dementia [27], outperforming several conventional MRI measures.

The method involved several quantification parameters derived from the analysis. First, we demonstrated a clear relationship between mass density variation and refractive index fluctuations in the Brain TMA samples, resulting in a transmission intensity that linearly depends on mass density. These variations are mostly related to mass density, which is connected to the fractal dimension of biological tissue samples as 2D fractals. Our explicit demonstration that mass density variations modulate refractive index, and therefore transmission intensity, provides a physical link between histological structure and the optical signal that underpins fractal quantification.

Now, the fractal dimension estimation also varies with the threshold. We demonstrated the threshold dependence of fractal dimension. Interestingly, the fractal dimension emerges as a function of the grayscale threshold intensity. A novel approach arises from this, as it can distinguish between these two groups and provide the optimal changes in fractal dimension values at the optimal threshold. Therefore, this fractal dimension acts as a biomarker for disease diagnosis.

Then, we analyzed the distribution of fractal dimension, which becomes a long-tailed distribution, and the fractal parameters *Mean($D_f$)* and *STD($D_f$)* systematically increase with disease progression. Multifractal analysis showed minor changes in the multifractal parameters. To further improve quantification, we used fractal functional transformation. However, we aimed to obtain a Gaussian distribution of the fractal functional transformation, as it is easier to interpret statistically. The distribution parameters, *Mean(ln($D_{tf}$))* and *STD(ln($D_{tf}$))*, show significant percentage changes from control to AD disease tissues. While previous AD studies have reported on distributions of $D_f$ or multifractal exponents [28,29], the explicit Gaussianization of fractal distributions and their use as potential biomarkers appears to be a novel contribution of this work and may facilitate integration into statistical and machine-learning classifiers.

To support our diagnosis process, our well-known light localization technique, IPR analysis, was implemented, which showed an enhancement in detecting structural disorder at the nanoscale [11,20,30]. The IPR is sensitive to refractive-index heterogeneity at subwavelength scales, making it particularly useful for examining early pathological changes. We observed that the IPR parameters, Mean IPR and STD of IPR, increase with disease progression, consistent with growing structural disorder in AD [31]. IPR values are linearly related to the strength of structural disorder strength indicating that increases in sample size and mass density fluctuations amplify disorder signatures. These findings position IPR-derived parameters as strong candidates for nanoscale biomarkers of neurodegeneration.

Taken together, the convergence of fractal, multifractal, fractal functional transformation, and IPR metrics provides a rigorous, multi-scale quantification framework for assessing tissue alterations in AD. Each analytical component captures complementary aspects of structural organization from global self-similarity and regional heterogeneity to nanoscale refractive-index disorder. The alignment of these independent measures reinforces the robustness of the framework and highlights the value of integrating physics-inspired methods into neuropathological analysis. Collectively, our results demonstrate that nanoscale structural disorder in AD can be quantified through a cohesive set of interpretable numerical biomarkers derived from standard bright-field microscopy images. This multi-parametric approach strengthens the potential for early detection, improves disease staging, and enhances objectivity in clinical pathology. Future integration with machine learning and larger patient cohorts may further refine diagnostic performance and accelerate clinical translation.

———————

**Acknowledgments:** We thank the Mississippi State and UTHSC imaging facilities for imaging.

**Funding:** This work was partially supported by the National Institutes of Health (NIH) under grant number R21 CA260147 to PP, and MMK's work was supported by NIH grant R03AG075597 and Alzheimer's Association grant AARG-NTF-22-972518.

# References

[1] H. W. Querfurth and F. M. LaFerla, Alzheimer's Disease, N Engl J Med **362**, 329 (2010).
[2] 2024 Alzheimer's disease facts and figures, Alzheimer's & Dementia **20**, 3708 (2024).
[3] E. T. Ziukelis, E. Mak, M.-E. Dounavi, L. Su, and J. T O'Brien, Fractal dimension of the brain in neurodegenerative disease and dementia: A systematic review, Ageing Research Reviews **79**, 101651 (2022).
[4] C. Sarkar, D. Gupta, M. Singh, A. Mahapatra, D. Jain, and M. Sharma, Comparative analysis of diagnostic accuracy of different brain biopsy procedures, Neurol India **54**, 394 (2006).
[5] J. G. Elmore et al., Diagnostic Concordance Among Pathologists Interpreting Breast Biopsy Specimens, JAMA **313**, 1122 (2015).
[6] L. Elkington, P. Adhikari, and P. Pradhan, Fractal Dimension Analysis to Detect the Progress of Cancer Using Transmission Optical Microscopy, Biophysica **2**, 1 (2022).
[7] L. G. Da Silva, W. R. S. Da Silva Monteiro, T. M. De Aguiar Moreira, M. A. E. Rabelo, E. A. C. P. De Assis, and G. T. De Souza, Fractal dimension analysis as an easy computational approach to improve breast cancer histopathological diagnosis, Appl. Microsc. **51**, 6 (2021).
[8] S. Maity, M. Alrubayan, I. Apachigwao, D. Solanki, and P. Pradhan, *Optical Probing of Fractal and Multifractal Connection to Structural Disorder in Weakly Optical Disordered Media: Application to Cancer Detection*.
[9] S. Maity, M. Alrubayan, and P. Pradhan, *Decoding Breast Cancer in X-Ray Mammograms: A Multi-Parameter Approach Using Fractals, Multifractals, and Structural Disorder Analysis*.
[10] I. Apachigawo, D. Solanki, R. Tate, H. Singh, M. M. Khan, and P. Pradhan, Fractal Dimension Analyses to Detect Alzheimer's and Parkinson's Diseases Using Their Thin Brain Tissue Samples via Transmission Optical Microscopy, Biophysica **3**, 569 (2023).
[11] P. Pradhan, D. Damania, H. M. Joshi, V. Turzhitsky, H. Subramanian, H. K. Roy, A. Taflove, V. P. Dravid, and V. Backman, Quantification of nanoscale density fluctuations using electron microscopy: Light-localization properties of biological cells, Applied Physics Letters **97**, 243704 (2010).
[12] R. D. King, B. Brown, M. Hwang, T. Jeon, and A. T. George, Fractal dimension analysis of the cortical ribbon in mild Alzheimer's disease, NeuroImage **53**, 471 (2010).
[13] S. Bhandari, S. Choudannavar, E. R. Avery, P. Sahay, and P. Pradhan, Detection of colon cancer stages via fractal dimension analysis of optical transmission imaging of tissue microarrays (TMA), Biomed. Phys. Eng. Express **4**, 065020 (2018).
[14] H. F. Jelinek, N. T. Milošević, A. Karperien, and B. Krstonošić, *Box-Counting and Multifractal Analysis in Neuronal and Glial Classification*, in *Advances in Intelligent Control Systems and Computer Science*, edited by L. Dumitrache, Vol. 187 (Springer Berlin Heidelberg, Berlin, Heidelberg, 2013), pp. 177–189.
[15] A. Chhabra and R. V. Jensen, Direct determination of the f(α) singularity spectrum, Phys. Rev. Lett. **62**, 1327 (1989).
[16] A. Barbora, S. Karri, M. A. Firer, and R. Minnes, Multifractal analysis of cellular ATR-FTIR spectrum as a method for identifying and quantifying cancer cell metastatic levels, Sci Rep **13**, 18935 (2023).
[17] Y. M. Elgammal, M. A. Zahran, and M. M. Abdelsalam, A new strategy for the early detection of alzheimer disease stages using multifractal geometry analysis based on K-Nearest Neighbor algorithm, Sci Rep **12**, 22381 (2022).
[18] E. Amin, Y. M. Elgammal, M. A. Zahran, and M. M. Abdelsalam, Alzheimer's disease: new insight in assessing of amyloid plaques morphologies using multifractal geometry based on Naive Bayes optimized by random forest algorithm, Sci Rep **13**, 18568 (2023).
[19] A. J. Joseph and P. N. Pournami, Multifractal theory based breast tissue characterization for early detection of breast cancer, Chaos, Solitons & Fractals **152**, 111301 (2021).
[20] P. Pradhan, D. Damania, H. M. Joshi, V. Turzhitsky, H. Subramanian, H. K. Roy, A. Taflove, V. P. Dravid, and V. Backman, Quantification of nanoscale density fluctuations by electron microscopy: probing cellular alterations in early carcinogenesis, Phys. Biol. **8**, 026012 (2011).

[21] P. Sahay, H. M. Almabadi, H. M. Ghimire, O. Skalli, and P. Pradhan, Light localization properties of weakly disordered optical media using confocal microscopy: application to cancer detection, Opt. Express **25**, 15428 (2017).

[22] P. Adhikari, P. K. Shukla, F. Alharthi, R. Rao, and P. Pradhan, Photonic technique to study the effects of probiotics on chronic alcoholic brain cells by quantifying their molecular specific structural alterations via confocal imaging, Journal of Biophotonics **15**, e202100247 (2022).

[23] P. Sahay, A. Ganju, H. M. Almabadi, H. M. Ghimire, M. M. Yallapu, O. Skalli, M. Jaggi, S. C. Chauhan, and P. Pradhan, Quantification of photonic localization properties of targeted nuclear mass density variations: Application in cancer-stage detection, Journal of Biophotonics **11**, e201700257 (2018).

[24] P. Pradhan and S. Sridhar, Correlations due to Localization in Quantum Eigenfunctions of Disordered Microwave Cavities, Phys. Rev. Lett. **85**, 2360 (2000).

[25] J. Fröhlich, F. Martinelli, E. Scoppola, and T. Spencer, Constructive proof of localization in the Anderson tight binding model, Commun.Math. Phys. **101**, 21 (1985).

[26] W. M. C. Foulkes and R. Haydock, Tight-binding models and density-functional theory, Phys. Rev. B **39**, 12520 (1989).

[27] C. Marzi, R. Scheda, E. Salvadori, A. Giorgio, N. De Stefano, A. Poggesi, D. Inzitari, L. Pantoni, M. Mascalchi, and S. Diciotti, Fractal dimension of the cortical gray matter outweighs other brain MRI features as a predictor of transition to dementia in patients with mild cognitive impairment and leukoaraiosis, Front Hum Neurosci **17**, 1231513 (2023).

[28] S. Lahmiri and M. Boukadoum, Alzheimer's disease detection in brain magnetic resonance images using multiscale fractal analysis, ISRN Radiol **2013**, 627303 (2013).

[29] T. Azizi, The Fractal Geometry of Alzheimer's disease, Alzheimer's & Dementia **18**, e060395 (2022).

[30] P. Adhikari, M. Hasan, V. Sridhar, D. Roy, and P. Pradhan, Studying nanoscale structural alterations in cancer cells to evaluate ovarian cancer drug treatment, using transmission electron microscopy imaging, Phys Biol **17**, 036005 (2020).

[31] F. Alharthi, I. Apachigawo, D. Solanki, S. Khan, H. Singh, M. M. Khan, and P. Pradhan, Dual Photonics Probing of Nano- to Submicron-Scale Structural Alterations in Human Brain Tissues/Cells and Chromatin/DNA with the Progression of Alzheimer's Disease, Int J Mol Sci **25**, 12211 (2024).